\documentclass{article}

\usepackage[english]{babel}

\usepackage[letterpaper,top=2cm,bottom=2cm,left=3cm,right=3cm,marginparwidth=1.75cm]{geometry}

\usepackage{amssymb}
\usepackage{latexsym}

\usepackage{natbib}
\usepackage{authblk}

\usepackage{url}
\usepackage{xcolor}
\definecolor{newcolor}{rgb}{.8,.349,.1}
\usepackage{graphicx}
\usepackage[colorlinks=true, allcolors=blue]{hyperref}

\title{The characteristics of flare- and CME-productive solar active regions}
\author[1]{Ioannis Kontogiannis}
\affil[1]{Leibniz-Institut f\"{u}r Astrophysik Potsdam (AIP), An der Sternwarte 16 14482 Potsdam, Germany}
\begin{document}
\maketitle

\begin{abstract}
Solar flares and coronal mass ejections (CMEs) cause immediate and adverse effects on the interplanetary space and geospace. In an era of space-based technical civilization, the deeper understanding of the mechanisms that produce them and the construction of efficient prediction schemes are of paramount importance. The source regions of flares and CMEs exhibit some common morphological characteristics, such as $\delta$-spots, filaments and sigmoids, which are associated with strongly sheared magnetic polarity inversion lines, indicative of the complex magnetic configurations that store huge amounts of free magnetic energy and helicity. The challenge is to transform this empirical knowledge into parameters/predictors that can help us distinguish efficiently between quiet, flare-, and CME-productive (eruptive) active regions. This paper reviews these efforts to parameterize the characteristics of eruptive active regions as well as the importance of transforming new knowledge into more efficient predictors and including new types of data. Magnetic properties of active regions were first introduced when systematic ground-based observations of the photospheric magnetic field became possible and the relevant research was boosted by the provision of near real time, uninterrupted, high-quality observations from space, which allowed the study of large, statistically significant samples. Nonetheless, flare and CME prediction still faces a number of challenges. The magnetic field information is still constrained at the photospheric level and accessed only from one vantage point of observation, thus there is always need for better predictors; the dynamic behavior of active regions is still not fully incorporated into predictions; the inherent stochasticity of flares and CMEs renders their prediction probabilistic, thus benchmark sets are necessary to optimize and validate predictions. To meet these challenges, researchers have put forward new magnetic properties, which describe different aspects of magnetic energy storage mechanisms in active regions and offer the opportunity of parametric studies for over an entire solar cycle. This inventory of features/predictors is now expanded to include information from flow fields, transition region and coronal spectroscopy, data-driven modeling of the coronal magnetic field, as well as parameterizations of dynamic effects from time series. Further work towards these directions may help alleviate the current limitations in observing the magnetic field of higher atmospheric layers. In this task, fundamental and operational research converge, with promising results which could stimulate the development of new missions and lay the ground for future exploratory studies, also profiting from and utilizing the long anticipated observations of the new generation of instruments.
\end{abstract}

\section{Introduction}
\label{intro}
Solar flares are sudden, localized brightenings of the solar atmosphere evident throughout the entire electromagnetic spectrum \citep{Fletcher_etal11,2011LRSP....8....6S}. They can release energy in excess of 10$^{25}$\,J, within minutes, via both thermal and non-thermal processes. Coronal mass ejections (CMEs) are expulsions of solar coronal plasma into the interplanetary space with velocities that can be as high as 3500\,kms$^{-1}$ and mass and energy of the order of 10$^{12}$\,kg and 10$^{25}$\,J, respectively \citep{chen11,webb12}. Flares are categorized in a logarithmic scale according to their soft X-ray emission in the 1 – 8\,\AA\ range, as recorded by the Geostationary Operational Environmental Satellites (GOES). From strongest to weakest these classes are X, M, C, B, and A, complemented by decimal subclasses (M1.0, C5.2, etc.). Flares of M1.0 and higher are often termed major flares, due to their severe impact on space weather. The association between flares and CMEs is not trivial. Statistically, the percentage of flares associated with CMEs increases with the strength of the flare, with the strongest X-class flares almost always associated with a CME \citep{2005JGRA..11012S05Y}. There are, however, differences in the reported percentages, depending on the solar cycle and the instrument/method of detection \citep{2019SSRv..215...39L}.  
A flare associated with a CME is called eruptive, otherwise, it is a confined flare. The term solar eruption usually describes eruptive flares and the associated CMEs. Both flares and CMEs are facets of the solar activity, usually originating from strong, extended, and often complex magnetic structures called active regions.

Flares and CMEs can affect the space near Earth in many ways, and these effects can occur within minutes, hours and/or days. The enhanced X-ray and EUV emission affect immediately the electron density in a range of ionospheric altitudes, disrupting radio communications, while the subsequent expansion of the atmosphere increases the drag on low-altitude satellites. The direct electromagnetic emission and the energetic particles, which are accelerated in situ and/or by CME fronts are hazardous to space-borne instrumentation and crew and can severely damage infrastructure (space-borne and ground-based), disrupting communications, navigation systems, and power grids \citep[see e.g.,][]{2009RaSc...44.0A17T,2016AdSpR..57..418K,2019RSPTA.37780098B,2019RSPTA.37780097D}.

Understanding these phenomena and mitigating their consequences on space-borne and ground-based infrastructures translates to understanding the mechanisms behind flares and CMEs and to developing efficient prediction schemes. Although the fundamental physics is arguably known \citep{2002A&ARv..10..313P}, the specific mechanisms still lack understanding and our ability to predict solar eruptions remains limited. Part of the current research has focused on the evolution of active regions, their characterization with respect to flaring and CME activity and the study of specific mechanisms that lead to these eruptive phenomena.

Active regions store enormous amounts of magnetic energy, which is transformed to the thermal and non-thermal components of solar eruptions via complex magnetic interactions and is more than enough to power solar eruptions. Both for understanding purposes as well as for prediction, the pre-eruptive evolution of active regions is parameterized through properties derived from photospheric vector and/or line-of-sight magnetograms. Some of these properties can be proxies of physical quantities, such as the total unsigned magnetic flux, the vertical current density, the free energy and the helicity. Others characterize the compactness and complexity of active regions, emphasize the presence of strong magnetic field gradients and polarity inversion lines (PILs) or distil information relevant to physical processes such as emergence of magnetic flux and/or shearing motions. The statistical association of these (mostly but not always) magnetic properties with eruptivity has been demonstrated in various samples of flaring and non-flaring regions and their efficiency (or performance) varies per study and sample.

The relevant research flourished with the launch of the Solar Dynamics Observatory \citep[SDO;][]{pesnell_etal12}, when a flow of near real-time constant quality photospheric vector magnetograms of active regions was provided by the Helioseismic and Magnetic Imager \citep[HMI; ][]{scherrer12}. To further support research and operation, the HMI team issues the Space Weather HMI Active Region Patches (SHARPs), cut-outs of vector magnetograms of regions of interest (such as active regions), along with a set of predictors suitable for space-weather research \citep{bobra14}. This type of data enabled the study of large, statistically significant samples with machine learning (ML) and thus advanced the prediction of flares, the verification of forecasting schemes and the ranking of magnetic properties of the source regions \citep[see e.g.,][to name but a few]{Bobra_couvidat15,bobra_ilonidis16,Florios:18}. 

Several reviews already exist regarding the properties of flare and CME-productive active regions and the mechanisms that lead to eruption, both from a theoretical and an observational point of view \citep{2009AdSpR..43..739S,Green18,Toriumi_Wang19,2020SSRv..216..131P}. The aim of this review is to present the progress made so far in quantifying the eruptive potential of active regions. This task largely consists in the transformation of a vague notion of complexity into classes and parameters; it was initially based on white light observations, but during the past two decades has relied mostly on photospheric magnetograms, while recently a plethora of EUV observations has also ignited interest in this type of data. More sophisticated parameters are being constructed, aiming to include information relevant to the upper atmosphere and temporal evolution. It is a field of research at the interface between fundamental and operational research, since the derived quantities are crucial for forecasting services, but can also offer insight into the mechanisms that dictate the pre-eruptive evolution of active regions. It is, therefore, hoped that the works presented here will provide a useful background and also inspire new methodologies, in this era of new solar observations. The reader can find complementary information regarding other crucial aspects of the methodology used in operational flare forecasting in \citet{Florios:18, leka_etal19a,leka_etal19b,park_etal20}. 

The paper is structured as follows: In Section~\ref{white_light} the characterization of active regions based on ground based observations is discussed, with emphasis on white-light observations. Section~\ref{mag_params} reviews the magnetic properties of flaring and erupting active regions, which largely quantify their non-potentiality. After a brief presentation of fractal and multifractal measures in Section~\ref{fractal}, the paper then provides a detailed account of recent efforts to produce new properties, mostly, but not exclusively, magnetic. Section~\ref{cmes} is dedicated to efforts on pinpointing those properties that are most suitable for CME prediction, while Section~\ref{time_series} reviews research on quantifying the temporal evolution of active regions and how this information could be incorporated in their prediction. Section~\ref{summary} summarizes and discusses the progress on characterizing flaring and erupting active regions. 

\begin{figure*}
\centering
\includegraphics[width=16cm]{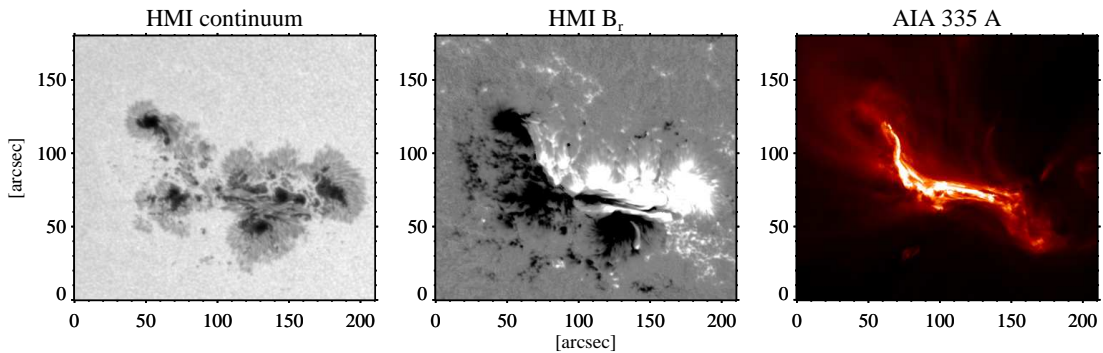}
\caption{Active region NOAA 11429 as seen on 6 March 2012. From left to right are the continuum at the Fe\,I\,6173\,\AA, recorded by the HMI, the corresponding map of the radial component of the photospheric magnetic field in the local frame of reference and the intensity at 335\,\AA\, recorded by the Atmospheric Imaging Assembly onboard the SDO. This region presented some of the most typical characteristics of highly-eruptive active regions, i.e., $\delta$-spot configuration, strong shearing motions, and a sigmoid seen in coronal EUV/Soft X-ray images. Courtesy of NASA/SDO and the AIA and HMI science teams.}
\label{fig:fig1}
\end{figure*}

\section{Quantifying the white-light morphology of active regions}
\label{white_light}

Figure~\ref{fig:fig1} shows a snapshot of active region NOAA 11429, one of the most flare- and CME-productive regions of Solar Cycle 24. It appeared already well-developed on the visible solar limb on 2 March 2012 and produced many flares as it traversed the solar disk. Most notable events were the two consecutive X-class flares produced during the first hours of March 7. As is most often the case with X-class flares, they were associated with fast CMEs, with pronounced effects on the geospace \citep{2016ApJ...817...14P}. Active region NOAA 11429 exhibited clearly some of the most typical characteristics of flare-productive regions: it had an anti-Hale orientation, i.e., the leading polarity was the positive one, contrary to what expected for the northern hemisphere during that solar cycle; it was exhibiting strong shear along the PIL; in EUV images of the corona it showed sigmoid structures, indicative of a pre-eruptive flux rope \citep{2015ApJ...809...34C}. 

Active region NOAA 11429 exhibited also another typical characteristic of flare-prolific regions, that is a $\delta-$spot, i.e., well-developed, opposite polarity sunspots sharing the same penumbra. This descriptor was introduced by \citet{1965AN....288..177K} and it is now used to complement the Mount Wilson (or Hale) classification of sunspots. The original Mount Wilson classification scheme was put forward by Hale \citep{1919ApJ....49..153H} and it was based on optical, ground-based, white-light and spectropolarimetric observations of active regions. The classes of Mount Wilson categorize sunspot groups in terms of the apparent distribution of their magnetic polarities. Four categories, $\alpha$, $\beta$, $\beta\gamma$ and $\gamma$ represent unipolar, bipolar, complex bipolar and irregular/multipolar regions. If the sunspot group contains a $\delta$-spot (which, by definition, cannot happen for an $\alpha$-type) it can be a $\beta\delta$, $\gamma\delta$ or $\beta\gamma\delta$, depending on the overall morphology of the active region. 

Although based on ground-based observations, the Mount Wilson classification requires information on the magnetic polarity of the sunspots. However, after decades of regular monitoring of the Sun, the white-light-only morphology of the most eruptive active regions had also been noted. These observations are the easiest and the longest, practically available since the era of Galileo and the first scholars, who used the telescope for solar observations, initially in the form of drawings and then, with the development of photography as photographic plates. The benefits of such observations are further appreciated today, since they span many solar cycles and facilitate the association of the number of sunspots and morphology with other aspects of solar activity and observed effects. Automated forecasting systems are based on categorization of these morphological characteristics and parameterizations, derived from large sample. The McIntosh classification system \citep{mcintosh90} offered such a basis and its development was already dictated by the need to improve the statistical association of flare occurrences with discrete classes of active region complexity.

Initially, the classification scheme of Zurich comprised nine classes of white-light complexity, denoted with letters from A to J (I omitted) based on typical evolutionary aspects of the largest sunspot groups \citep{1901ApJ....13..260C,1938ZA.....16..276W}. This system was modified to introduce two more classification criteria, namely the morphology of the penumbra of the largest sunspot (indicated by the letters x,r,s,a,h, and k) and the distribution of sunspots within the group (indicated by the letters x,o,i,c). Additionally, it was realized that some of the original Zurich classes (G and J) were redundant in the new scheme, thus leading to an expansion of the remaining seven out of the original Zurich classes into sixty. Although lacking magnetic field information, this classification is more detailed than the Mount Wilson scheme, also incorporating evolutionary aspects, which are missing from the latter. For example, classes at the extremes of the McIntosh scale, such as Axx or Hrx may correspond to $\alpha$-type, Bxo or Dsi can be $\beta$, while the compact and complex classes, like $Ekc$, $Eki$, $Fkc$, $Eki$ correspond to Mount Wilson types with the $\delta$-identifier. In fact, the latter categories are associated with the highest M- and X-class flaring rates, as demonstrated in \citet{mcintosh90}.

The \citet{mcintosh90} study showed the importance of context observations and large statistical samples in establishing the crucial characteristics of flare-productive regions. To this end, the digitization of such observations may expand this sample of observations as far back as the $18^{th}$ century \citep[See][for a non-exhaustive but informative review of digitization efforts]{2020AN....341..575P}. Moreover, the study also showed that the association between flare-productivity and active region characteristics is statistical: even though it is clear that the most complex classes are associated with higher flaring rates, flares may occur in simpler regions irrespective of the phase of their evolution. This statistical association effectively renders the prediction of flares probabilistic, and transforming probabilities to binary forecasts requires carefully designed validation strategies, as demonstrated in \citet{bloomfield12}. The probabilistic nature of flare and CME prediction has persisted and is not expected to change. What has, in fact, changed so far are the data sources, the sophistication of quantifying morphological and magnetic complexity and an increasing trend of reliance on ML for the prediction models.


\section{Magnetic field parameters}
\label{mag_params}

The increasing access to good quality photospheric magnetograms during the 1990's, brought about by the development of ground-based and space-borne instruments facilitated the detailed studies of the magnetism in active regions and its evolution. Exploratory studies were establishing the important role of complexity, size, and rapid emergence of magnetic flux or cancellation in flare-productive regions. Additionally, it was possible to measure physical magnitudes and proxies thereof, i.e., quantities representative of these mechanisms and of the accumulation of magnetic energy in active regions. 

A series of landmark papers by \citet{2003ApJ...595.1277L,2003ApJ...595.1296L,2006ApJ...646.1303B,2007ApJ...656.1173L} illustrated the potential of using such magnetic field-related quantities in solar flare prediction. They used photospheric vector magnetograms acquired by the University of Hawai'i Imaging Vector Magnetograph, from which they calculated several magnetic parameters. The most rudimentary is related to the magnitude of the magnetic field vector and/or its components, measured at each pixel. Since the magnetic field flux density is a signed quantity, its net and total unsigned value can be calculated, as well as its sign imbalance. Vector magnetograms can be used to calculate the inclination angle of the field at the photosphere. 

The current instrumentation allows the determination of the magnetic field only at the photosphere, which imposes limitations to the physical magnitudes that can be derived. It is possible to determine the horizontal gradient of the magnetic field, which is a proxy of the compactness of the active region and, thus, \citet{2003ApJ...595.1277L} argue that it quantifies the sunspot distribution component of the McIntosh classification. In a similar fashion, one can also calculate (only) the vertical component of the electric current density, which can be treated similarly to the magnetic field density, thus determining the total (net) and the total unsigned electric current, as well as the total value per sign and per magnetic polarity. These parameters are proxies of the energy stored in the current-carrying magnetic field and the net currents injected into the corona. 

The product of the vertical magnetic field density and the vertical electric current, $J_{z}B_{z}$, is used as a proxy of the current helicity, whereas the ratio between the vertical current density and the vertical magnetic flux density is a proxy of the twist of the magnetic field. The twist can also be determined by using the vertical component of the magnetic field as a boundary condition and calculating the linear force-free magnetic field at the photosphere \citep[see e.g.,][]{1981A&A...100..197A}, thus producing one twist value representative for the entire field of view. Similarly, performing an electric current-free (potential) field extrapolation, a proxy of the free magnetic energy density can be derived as the square of the difference between the observed and the potential magnetic fields. It is noted here in passing that a reliable determination of the free magnetic energy is sensitive to measurement errors, to the flux imbalance of the studied region (if any) and the assumptions made regarding the force free magnetic field \citep[see e.g.,][]{2007ApJ...671.1034G,2012ApJ...759....1G}. Additionally, the angle between the two magnetic field vectors is a proxy of the shear angle, which can be calculated on a pixel-by-pixel basis. From this distribution of values, the total/average value for the entire FOV or for regions of interest (such as along PILs) can be determined. Another parameter implemented in these studies was the length of the neutral line (or PIL) for which the shear angle was higher than a specific value (e.g., 45$^{\circ}$). 

For the quantities that were calculated on a pixel-by-pixel basis, in addition to averages, net values and unsigned totals, the higher-order moments of their distributions were also used, namely, the standard deviation, skewness and kurtosis. It was argued that all flare-related mechanisms, e.g., flux emergence, flux cancellation, shearing and twisting (which could lead to an increase in the horizontal components) would produce an observable effect on the related quantities. 

The predictors described above were introduced and studied for a limited sample of events and active regions in \citet{2003ApJ...595.1277L}. The same set of predictors was then extracted from a larger set of observations and discriminant analysis was performed to examine the efficiency of these parameters in discriminating flaring from non-flaring regions \citep{2003ApJ...595.1296L}. 

The characteristic of these quantities is that they are derived from photospheric magnetograms and they are related strictly to the magnetic field state at the photosphere. However, we know that the processes that lead to flares and CMEs take place above the chromosphere. To include quantities relevant to the connectivity and the state of the magnetic field in the corona \citet{2006ApJ...646.1303B} employed a magnetic charge topology model \citep{2005ApJ...629..561B}. Using a flux tesselation method, the photospheric magnetograms were segmented into non-overlapping unipolar partitions, representing the traces of magnetic charges. This ensemble of partitions was then used to determine a series of topological quantities, such as the number of null points, the separators and their length, the connectivity matrix, the magnetic flux associated with the connections, the magnetostatic energy of the collection of magnetic charges, etc. Discriminant analysis on a small set of events showed the prospect of using MCT-related parameters and in the last paper of the series \citep{2007ApJ...656.1173L}, the array of parameters derived from the photospheric vector magnetograms was tested on a large sample of events. 

\begin{figure*}
\centering
\includegraphics[width=16cm]{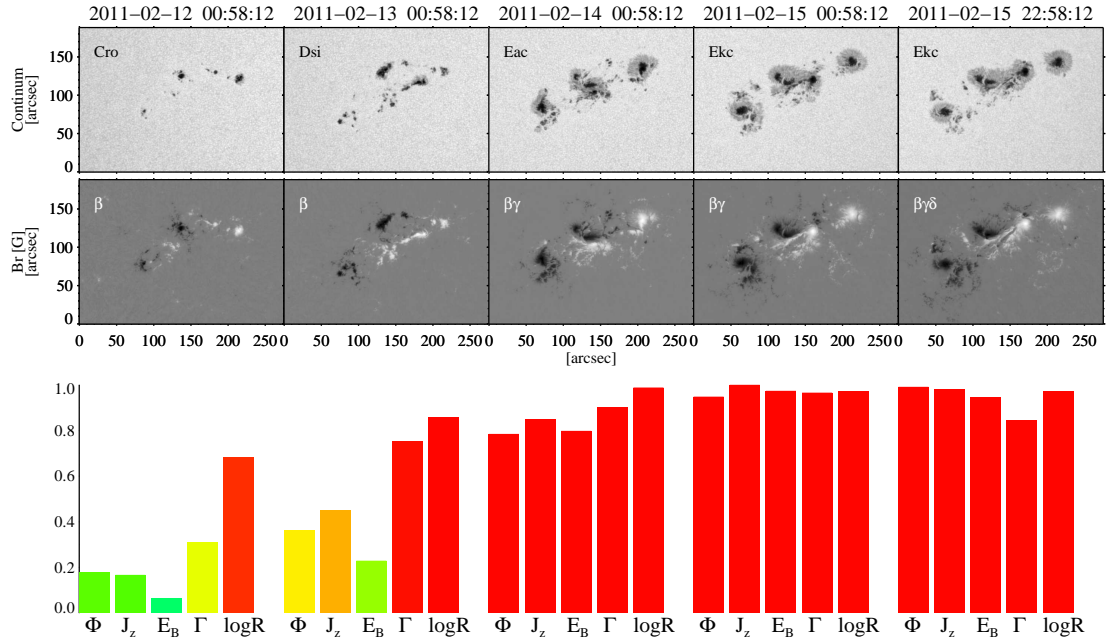}
\caption{The evolution of NOAA 11158 in McIntosh and Mount Wilson classes as well as in terms of the instantaneous values of some magnetic parameters. The chosen parameters were the total unsigned magnetic flux,$\Phi$, the vertical component of the total unsigned electric current density, $J_{z}$, the total photospheric magnetic free energy density, $E_{B}$, the mean shear angle, $\Gamma$, and Schrijver's $R$. All parameteres are included in the SHARP data product and for the purposes of this figure they have been scaled between their minimum and maximum value over the interval between 2011-02-10 21:58 and 2011-02-15 22:58. Courtesy of NASA/SDO and the AIA and HMI science teams. McIntosh and Mount Wilson classes were taken from \url{solarmonitor.org}.}
\label{fig:fig2}
\end{figure*}

Other studies focused on specific aspects of the magnetic field non-potentiality. Building on the knowledge that CMEs are produced preferentially by regions with sigmoids, i.e., S-shaped structures seen in some EUV wavelengths and soft X-rays, and that these structures partly outline strong and highly sheared PILs, \citet{2001JGR...10625185F} examined the associated CME productivity with two related parameters extracted from vector magnetograms. They calculated the length of strong-field, strong-shear, main neutral line, $L_{SS}$, and the net vertical electric current, $I_{N}$, as measures of the global non-potentiality of the studied regions. In a subsequent work \citep{2002ApJ...569.1016F} they included also the twist parameter, i.e., the ratio between the vertical electric current and magnetic flux densities. In order to facilitate the use of the line-of-sight magnetograms recorded from both the ground and by the Michelson Doppler Imager \citep[MDI;][]{scherrer95} onboard the Solar and Heliospheric Observatory, they developed a proxy for the PIL that could be derived only from the line-of-sight component of the photospheric magnetic field, called $L_{SG}$ \citep{2003JGRA..108.1380F}. This was then calculated for an extended set of ground-based magnetograms and it was compared with other non-potentiality measures derived from vector magnetograms \citep{2006ApJ...644.1258F}. For the PIL-related parameters introduced in these studies only the regions where the shear (for $L_{SS}$) or the horizontal gradient (for $L_{SG}$) were higher than specific thresholds were taken into account. For active regions with many PILs the same authors developed generalized versions of $L_{SS}$ and $L_{SG}$, namely the the $WL_{SS}$ and $WL_{SG}$, which were the gradient-weighted integral length of strong shear neutral lines calcualted from vector and LOS magnetograms \citep{2008ApJ...689.1433F}. These were found to be correlated with the free magnetic energy of active regions and thus they are now used as its proxies in the MAG4 flare-prediction service (\url{https://www.uah.edu/cspar/research/mag4-page/}). 

Having established that PILs are telltale signs of imminent intense flaring activity, several parameterizations were put forward around the same era. Based on the empirical fact that strong flares are observed in the vicinity of PILs,  \citet{schrijver07} introduced the quantity $R$, also called \textit{``Schrijver's R''}. The PILs are located using morphological image processing and R is the total unsigned magnetic flux in a 15\,Mm-wide region along the PIL. In principle the decimal logarithm of the quantity is being used ($logR$). For the examined sample of active regions, which spanned from 1999 to 2006, it was found that no major flares (M- or X-class) occurred for regions with $logR$ lower than 2.8, but when this value was higher than 5 the flaring probability reached unity. Another morphological quantity introduced to quantify the presence of strong, i.e., highly-sheared, PILs is the effective connected magnetic field strength, $B_{eff}$ \citep{georgoulis07}. This quantity is derived from the connectivity matrix of the partitioned magnetogram \citep{2005ApJ...629..561B}, and it is the total of the magnetic flux associated with all connections, weighted by their length. This quantity is used in an operational forecasting scheme, the Athens Effective Solar Flare Forecasting (A-EFFORT), which is part of the European Space Agency (ESA) Space Situation Awareness (SSA) Programme. Detection algorithms of PILs such as the one presented in \citet{2010ApJ...723..634M} can be used to determine more rudimentary characteristics such as the length of the main PIL, the total length of PILs in an active regions and the associated magnetic flux \citep[see also:][]{2018SoPh..293....9G}, which is similar to Schrijver's R. Another measure of the compactness of active regions, called Ising Energy, a term inspired by the Ising model of ferromagnetism in statistical mechanics, was introduced by \citet{ahmed10}. This metric is the sum of the inverse square distances between all possible opposite polarity pixel pairs of an active region. A higher number of closely neighboring, opposite polarity magnetic field concentrations, which is typically the case in the presence of strongly sheared PILs and $\delta$-spots, will lead to higher values of Ising energy.  

Parameters derived from the magnetic field still remain the main features for prediction models. The parameters included in \citep{2007ApJ...656.1173L}, along with Shcrijver's R and $WL_{SG}$ are included in the list of space weather related parameters of the SHARP data products \citep{bobra14}. This data set spans now more than one full solar cycle and it is invaluable for testing new forecasting schemes. Recently, this data product was extended to include the LOS photospheric magnetograms recorded by the MDI instrument. The Space Weather MDI Active Region Patches \citep[SMARPS;][]{2021ApJS..256...26B} effectively extend our inventory of some magnetic parameters over two solar cycles \citep[see also][for an assessment of using two solar cycle data in training prediction models]{2022arXiv220403710S}. It is also worth mentioning that in the context of the recent project Flare Likelihood And Region Eruption Forecasting \citep[FLARECAST;][]{2021JSWSC..11...39G} all available parameters were gathered and calculated for the SHARP cut-outs and comprise the data base of the project, upon which forecasting models were based. The list comprises 209 predictors and includes all the aforementioned quantities, along with those discussed in Section~\ref{fractal} as well as newly developed ones, described in Section~\ref{new_predictors}.

\begin{figure}
\centering
\includegraphics[width=12cm]{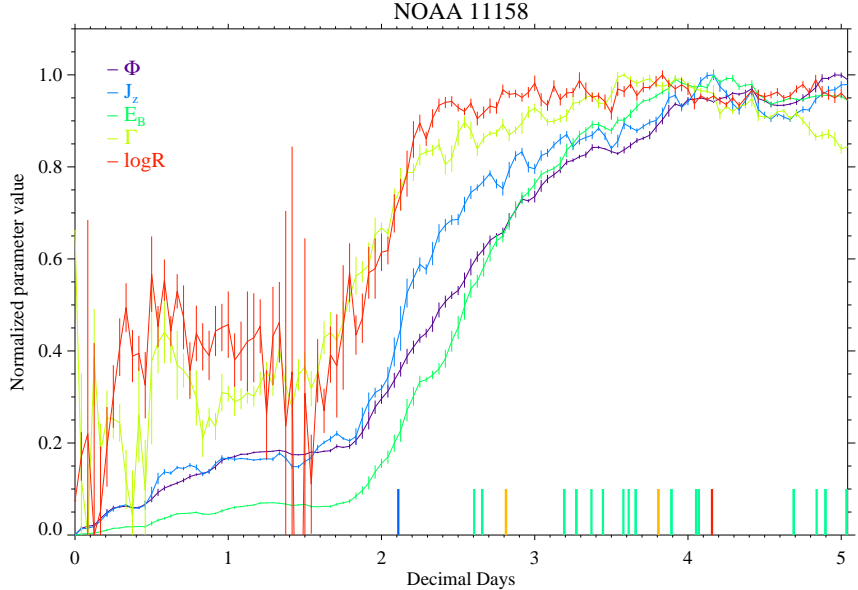}
\caption{Temporal evolution of the magnetic parameters shown in Figure~\ref{fig:fig2} for NOAA 11158, starting at 2011-02-10 21:58 UT. Blue, green, orange and red vertical lines at the lower part of the plot indicate B-, C-, M-, and X-class flares, correspondingly.}
\label{fig:fig2b}
\end{figure}

An example of how these magnetic parameters reflect the evolution of active regions is shown in Figures~\ref{fig:fig2} and \ref{fig:fig2b}. The top and middle rows of Figure~\ref{fig:fig2} show the evolution of the well-studied active region NOAA 11158, from 12 to 15 February 2011, in terms of overall white light morphology (along with McIntosh classes) and the photospheric longitudinal magnetic field. The region emerged towards the end of February 10 and within a few days it evolved from a relatively simple bipolar region into a $\delta$-spot, producing several C- and M-class flares, as well as the first X-class flare of Solar Cycle 24. At the bottom row of Figure~\ref{fig:fig2}, the respective morphological evolution is given in terms of five magnetic parameters, namely, the total unsigned magnetic flux,$\Phi$, the vertical component of the total unsigned electric current density, $J_{z}$, the total photospheric magnetic free energy density, $E_{B}$, the mean shear angle, $\Gamma$, and Schrijver's $R$. All parameters are provided with the SHARP data product for NOAA 11158 and here they have been normalized to their minimum (set to 0) and maximum (set to 1) values during the lifetime of this active region and are shown without their units. It is clear how the increase in complexity, both in terms of white-light morphology and magnetic field reflects on the values of all the magnetic parameters. The intensely flaring phase of the region, after February 13, followed the rapid increase of all parameters. In a statistical sense, higher values of these parameters should be associated with higher flaring rates, as is the case in this example.

However, the temporal evolution of the five parameters is not identical since each of them is associated with a slightly different aspect of the region's non-potentiality (size, appearance and development of PIL and shear, etc.). Figure~\ref{fig:fig2b} shows the corresponding time series in more detail. The time series have a cadence equal to 1\,h, instead of the 12\,min cadence originally provided with SHARPs, and are also normalized to their minimum and maximum values. As the active region emerged, the magnetic flux and free energy densities increased slowly during the first two days and then rapidly, as the region entered its intensely flaring phase. The electric current density followed a similar evolution, but with a more steep increase after the second day. The increase in the three parameters persisted until at least the end of the fifth day. The evolution of the mean shear and $logR$ exhibits some notable differences. These parameters are most closely associated with PILs, they exhibited some noisy behavior during the initial stages of emergence, when the first bipoles appeared and grew, then diminished as the polarities separated, but increased sharply before the end of the second day when the main PIL formed. The mean shear continued to increase until a few hours before the X-class flare and then decreased, while $logR$ was roughly constant during the entire flaring phase of the active region. These differences in evolution may be linked to the specific mechanisms that are responsible for intense flaring, such as strong converging and shearing motions, the development of magnetic shear, the emergence of magnetic flux and flux cancellation, and deserve further attention. In the context of flare prediction they can justify some of the differences regarding the efficiency of some parameters, the assessment of which depends on sample selection and prediction method.


\section{Fractal and multifractal properties}
\label{fractal}

The visual complexity of the photospheric magnetic field of active regions and their evolution towards critical states and destabilization have inspired their study in terms of their fractal and turbulent properties. In these cases the distribution of the magnetic field with values higher than a specific threshold is treated as an irregular shape whose fractal dimension can be determined. Since not only the spatial distribution, but also the values of the magnetic field may exhibit fractal properties, one can also examine the fractal properties of the distribution for varying indices and magnetic field thresholds, thus determining a multifractal spectrum. An account of early studies of the fractal properties of the magnetic field in quiet Sun and in plages, as well as a treatment of the distribution of the magnetic field flux density as a multifractal is given by \citet{1993ApJ...417..805L}. Since then, a number of studies have investigated the use of fractal measures as predictors of imminent flaring activity.

The simplest measure of the self-similarity of a set (in our case the magnetic field distribution of an active region) is the fractal dimension, which can be most simply determined using box counting. The process consists in covering the mask of the magnetic field distribution with a grid of boxes, counting the number of boxes containing at least one pixel of the mask, and repeating for various sizes of boxes. The number of boxes as a function of size should obey a power law, whose index is the fractal dimension. In principle, the fractal dimension of both the active region area and its boundary can be determined using the same method. \citet{2005ApJ...631..628M} applied the box-counting technique to a set of more than 9000 full-disk MDI magnetograms, and reported a tendency for flaring active regions to exhibit fractal dimension higher than 1.2.

The multifractal measures take into account the values of the magnetic field distribution. The corresponding methods result in the determination of a multifractal spectrum, from which specific indices may be used as predictors. The formalism to calculate the multifractal spectra of magnetic fields, in terms of generalized correlation dimensions and/or H\"{o}lder exponents and Hausdorff dimensions is described in \citet{1993ApJ...417..805L}. A series of studies have explored the connection between multifractal dimensions and flare productivity \citep[e.g.,][]{2005SoPh..228....5G,Conlon2008,2008SoPh..248..311H,2008SoPh..252..121I,2009A&A...506.1429C,2015SoPh..290..507G} and report a different behavior of the multifractal spectra and the values of specific multifractal dimensions of intensely flaring active regions. 

Another way to investigate the intermittency and turbulence present in a magnetic field distribution is through the multifractal structure function spectrum. \citet{2002ApJ...577..487A} studied the scaling behavior of the LOS magnetograms of four active regions with different flaring activity. The structure functions are equivalent to the correlation functions of a vector field (in this case the magnetic field) and are calculated as the statistical moments of the field increments, over a grid of separation vectors within the FOV. The range of separation vectors is defined as the inertial range. The structure functions follow power laws whose scaling indices can be determined through fitting. 

An assessment of the potential of fractal and multi-fractal parameters in flare prediction was attempted by \citet{2012SoPh..276..161G} and \citet{Georgoulis_2013}, for a representative sample of active regions observed by MDI and two case-studies of emerging active regions exhibiting different activity, observed by HMI, correspondingly. It was concluded that multi-scale parameters do not seem to provide a better chance of predicting flares than other fundamental or proxy parameters related to the non-potentiality of active regions. The latter study also highlights two important aspects of characterizing the non-potentiality of active regions, namely that it is possible to create more sophisticated proxy parameters to improve flare prediction and that the temporal evolution of the proxy and fundamental parameters need also be taken into account to this end. In Sections~\ref{new_predictors} and \ref{time_series}, recent efforts towards these two directions are reviewed. 


\section{Searching for new predictors}
\label{new_predictors}

The easy access to high-quality, near-real-time, photospheric magnetograms and context EUV coronal imaging for over a decade now, the accumulation of new and digitization of older synoptic observations, combined with the increasing concern about the implications of solar storms have further fueled research on improved prediction metrics and methods. One aspect of this research was the search for more efficient parameterizations of the pre-eruptive state of active regions, both in strictly magnetic and in non-magnetic terms. 

In a series of papers, \citet{korsos14,korsos15,korsos16} proposed a new way to measure the horizontal gradient of the magnetic field, based on white light observations. They argued that white-light images would suffer less than photospheric magnetograms from the geometric foreshortening effects, which affect observations away from the solar disk center and that these could be easily corrected using simple transforms. They used simultaneous observations of sunspot groups and photospheric magnetograms close to the solar disk center to calibrate the umbrae areas into magnetic field values. With this calibration relation at hand, they calculated the horizontal gradient of the magnetic field as the sum of all differences between the areas of opposite-polarity umbrae, weighted by the average separation distance between the leading and trailing polarities. Following their work on the photospheric horizontal magnetic gradient, they suggested its extension at higher atmospheric layers, with the aid of extrapolations of the magnetic field \citep{2020ApJ...896..119K}. The methodology put forward by \citet{korsos15} was applied to a sample of $\sim$10000 SHARP cut-outs by \citet{kontogiannis18} who also did a preliminary assessment of the efficiency of horizontal gradient in comparison with other new predictors. Additionally, combining elements from the methodology of \citet{korsos15} and \citet{2005ApJ...629..561B}, they introduced two new variations of the Ising energy \citep{ahmed10}, namely, the Ising energy of the magnetic partitions $E_{Ising,part}$ and the Ising energy of sunspot umbrae $E_{Ising,spot}$. These quantities were calculated in a similar manner as the original, but instead of opposite polarity pixel they considered opposite polarity partitions/umbrae. It was then demonstrated that both new quantities as well as the horizontal gradient, $G_{S}$, had increased potential as predictors.

Based on the method of calculating the amount of net currents injected to the corona suggested by \citet{geo_titov_mikic12}, two new predictors were introduced by \citet{kontogiannis17}, namely, the total unsigned non-neutralized electric current, $I_{NN,tot}$ and the maximum non-neutralized current $I_{NN,max}$. These were tested in a sample of active regions and also $\sim$10000 SHARP cut-outs and it was found that their values correlated well with the flare productivity of active regions and exhibited higher flaring rates than other commonly used electric current-related parameters. 

\begin{figure}
\centering
\includegraphics[width=12cm]{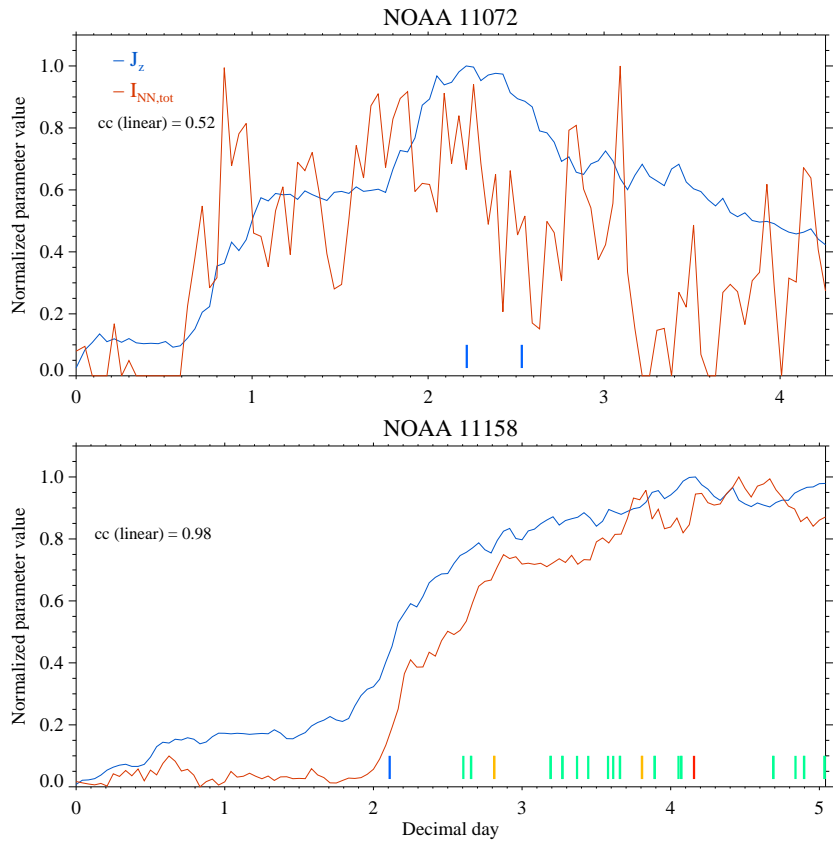}
\caption{The evolution of the vertical component of the total unsigned electric current density (blue) and the total unsigned non-neutralized electric current (red) for active regions NOAA 11072 (top) and 11158 (bottom), from the start of their emergence at 2010-05-20 16:22 and 2011-02-10 21:58\,UT correspondingly. The two time series are scaled to their maximum values to facilitate comparison. Blue, green, orange and red vertical lines indicate B-, C-, M-, and X-class flares, respectively.}
\label{fig:fig3}
\end{figure}

The difference between $I_{NN,tot}$ and the vertical component of the electric current density, $J_{z}$ is shown, for two active regions with different flaring activity, in Figure~\ref{fig:fig3}. Both parameters followed a similar trend as the active regions evolved. However, the values of the $I_{NN,tot}$ were lower than $J_{z}$ by roughly one order of magnitude (note that the time series in Figure~\ref{fig:fig3} were normalized to facilitate comparison), because the latter is calculated from the entire distribution of the magnetic field vector (within a bitmap mask) whereas the former only from the non-neutralized partitions. As per the proposed methodology, these are the magnetic partitions for which the electric current density is significantly higher than the erroneous numerical or error-propagated electric currents. These criteria, chosen by \citet{geo_titov_mikic12}, ensure that only neighboring partitions which have developed sheared PILs contribute to the value of $I_{NN,tot}$, whereas isolated magnetic partitions should, in principle, be neutralized and not contribute to the calculation. This important difference makes the relation between $J_{z}$ and $I_{NN,tot}$, non-trivial, since for simple active regions $I_{NN,tot}$ is zero while $J_{z}$ is not. As seen in Figure~\ref{fig:fig3}, the correlation coefficients between the two parameters calculated for the two active regions clearly differ. For NOAA 11072 the correlation is moderate (0.52) since most of the partitions of the region are neutralized. For NOAA 11158, the correlation is almost unity (0.98), since the region developed intense shearing motions as it evolved, leading to a similar increase in $I_{NN,tot}$. In fact, the strong correlation between the two parameters is due to the evolution observed after the second day of its emergence. Therefore, during this phase in the overall evolution of the active region the increase in $J_{z}$ is due to electric current injected to the corona by the non-neutralized partitions. On the contrary, this is not the case during the first two days: although $J_{z}$ increases monotonically following the growth of the region, this is not the case for $I_{NN,tot}$, as a result of low shear and absence of strong PILs during the early stage of the emergence. The association between non-neutralized electric currents and strongly sheared PILs is further discussed in \citet{geo_titov_mikic12} and \citet{kontogiannis17,kontogiannis19}, while more work is underway, but this example illustrates how exploratory research can lead to more refined predictors, i.e., parameters that reflect in more detail the specifics of the mechanisms that lead to flares and CMEs. 

Before discussing some alternative ways to utilize observations to derive predictors of flaring activity, it is worth mentioning that this research can also be aided by scrutiny of simulations. \citet{Guennou17} tested and developed some non-potentiality metrics and assessed their ability to differentiate between eruptive and non-eruptive numerical setups using numerical experiments of flux emergence instead of observations. Along with the importance of the parameters introduced by \citet{falconer08}, they also highlighted the efficiency of new parameters, such as the length of the strong-current MPIL and the corresponding integral of the magnetic field along it. 

Another aspect of active region evolution is relevant to the observed photospheric proper motions and flow fields. \citet{Welsch2009} used Fourier Local Correlation Tracking (FLCT) and the Differential Affine Velocity Estimator \citep[DAVE; ][]{2005ApJ...632L..53S} to extract the flow fields from MDI active region observations. Then they produced an array of quantities to parameterize the flow fields in order to investigate their association with flaring activity. The list included quantities related to the magnetic field, the flow field, their moments, the divergence and convergence close to the PIL, Schrijver's R, etc. They noted that their proxy of the Poynting flux, i.e., the sum of the product of flow velocity and square radial magnetic field was the one with the strongest association with flare flux. \citet{park18} used the Differential Affine Velocity Estimator for Vector Magnetograms \citep[DAVE4VM; ][]{2008ApJ...683.1134S} to measure the flow fields of a large sample of flaring and non-flaring active regions observed by HMI. Their focus was on the shear flows along the PIL and, thus, three quantities were constructed, namely, the mean, total (integral) and maximum along the PIL, as well as the corresponding horizontal and vertical flow speeds. Although all parameters show similar trends with flaring activity, their results suggest that a measure of the shear flows such as the integral along the MPIL could be used in flare prediction. The conclusion from these two works is that the flow field information in active region could be incorporated in the prediction of flares and CMEs, although more work is required.

The increasing abundance of high quality EUV context and spectroscopic observations of active regions in the UV have facilitated the study of sizeable samples. In this context, \citet{2020ApJ...891...17P} gathered Mg\,II\,h \& k lines spectra of active regions observed by the Interface Region Imaging Spectrograph \citep[IRIS][]{2014SoPh..289.2733D}. The spectra were represented in terms of ten spectral parameters (intensity, Doppler shift, line width, asymmetry, continuum, triplet emission at the red wing, k/h ratio, k3 intensity, peak ratios and peak separation), all of them reflecting physical conditions up to the chromosphere. They used both unsupervised - Principal Component Analysis (PCA) and t-distributed stochastic neighbor embeddings (t-SNE) - and supervised (neural networks) ML and concluded that the triplet emission, followed by intensity and total continuum were the leading parameters. Further utilization of non-magnetic information was done by \citet{2020arXiv201106433G}, who formulated parameters based on the Differential Emission Measure (DEM) of active regions, as reconstructed by AIA imaging. They used data provided by the Gaussian AIA DEM maps \citep[GAIA-DEM;][]{2012ApJS..203...25G,2012ApJS..203...26G} provided by the Multi Experiment Data \& Operation Center (MEDOC; \url{https://idoc-medoc.ias.u-psud.fr/}). The database contains maps of four parameters, namely, the emission measure (i.e. the temperature integral of the DEM curve), the maximum temperature and width of the DEM distribution (corresponding to the position of maximum and the width of the fitted Gaussian), as well as the $\chi^{2}$ of the fit. They demonstrated that the temporal derivative of the emission measure and the maximum temperature can be used to forecast flares, especially for shorter forecast windows. An additional interesting aspect of the aforementioned works on flow fields and the DEM-related parameters is that the proposed predictors also contain information on the dynamical evolution of the active regions.

During the past few years some different methods to parameterize active region complexity have been developed. \citet{2020JSWSC..10...13D} used topological analysis to construct the persistence diagrams of magnetograms using different magnetic flux thresholds and then transformed these into vectors, which they used as predictors to train ML algorithms. Additionally, they implemented geometrical features such as the number and size of positive (negative) elements, their average distance as well as the interaction factor, a measure similar to the Ising energy. They trained a neural network with the geometrical and topological features, complemented with the SHARP parameters and concluded that the new predictors could improve forecasts. \citet{2021SpWea..1902837S} also included features derived from topological analysis, such as variograms, along with their different versions of SHARP parameters. Since the classical SHARP parameters represent total or average values over a region around the active region, they produced modifications of these parameters by placing more weight on the region around the PIL, following the method of \citet{2020ApJ...892..140W}. Additionally, they employed spatial statistics tools such as Ripley's K function and variograms to quantify the clustering of pixels above specific magnetic flux density thresholds. Then, through a PCA they kept the top five components from each K function and topological features and concluded that the corresponding features perform equally well and sometimes better than the SHARP parameters.

\citet{2021ApJ...915...38C} parameterized PILs using a topological descriptor called $D$, which represents the number of PIL fragments in an active region. The opposite polarity fragments were detected using thresholding, segmentation and labeling of the LOS magnetograms. They used $D$ along with a version of Schrijver's $R$ and the properties extracted by the forecasting engine of FLARECAST \citep{2021JSWSC..11...39G}. They reported that $D$ was always included in the top-10 performing predictors in all realizations of the forecasting scheme. \citet{2022ApJ...927..156R} suggested a new metric called magnetic winding as a proxy of the magnetic helicity. The quantity was calculated without the need of three-dimensional extrapolations of the magnetic field and relied on the decomposition of the magnetic field into potential and current-carrying components. They presented five examples of flaring and non-flaring active regions to illustrate the potential of the new metric.

Finally, instead of calculating complexity parameters, abstract characteristics can be extracted from magnetograms. For instance, \citet{2017ApJ...834...11R} and \citet{2019ApJS..243...20A} used the Zernike moments of cut-outs of the active region magnetograms and EUV filtergrams as features for flare prediction.  Another way to quantify the complexity and eruptive potential of active regions is to directly feed magnetograms or filtergrams into ML algorithms \citep[see e.g.,][]{2019ApJ...884..175W,2018SoPh..293...48J,2021EP&S...73...64N,2021AdSpR..67.2544A}.


\section{Quantifying the potential to erupt: parameters suitable for CME prediction}
\label{cmes}

Although the standard solar model or eruptions \citep[also referred to as the CSHKP model, from the initials of][]{1964NASSP..50..451C, 1966Natur.211..695S,1974SoPh...34..323H,1976SoPh...50...85K} and its subsequent revisions describe both flares and CME's as aspects of a single eruptive process, the two phenomena are not always associated, as already discussed in the introduction. Although CMEs from active regions are always associated with flares, the opposite is not necessarily true. The study and prediction of flares have been prioritized for historical reasons, since they were discovered more than a century earlier than CMEs and for years they were considered to be the primary aspect of solar activity. This has now changed but there are also practical reasons for this ``bias'': while there is often enough time to model and predict the effects of CMEs upon their detection in white-light coronagraphic observations, the impact of flares is immediate. Nevertheless, the adverse effects of CMEs to geospace also call for accurate forecasting schemes. One way to accomplish this is to find which parameters are more effective for CME prediction, that is, which of them can help effectively distinguish eruptive from non-eruptive flares. An extra step is to modify some of the already existing parameters or even produce new ones, to improve their association with eruptions. Towards this direction, and given the fact that magnetic parameters are relevant mostly to the photospheric magnetic field, some recent studies have started to incorporate more elaborate information aiming to include the role of the constraining magnetic field.  

The phenomenology of solar eruptions is described by the standard model in terms of a core field, either a magnetic flux rope \citep{1999A&A...351..707T} or a sheared arcade \citep{1989ApJ...343..971V}, which is constrained by an overlying magnetic field. This two-dimensional model has been updated to accommodate observational aspects in three dimensions, such as the formation of sigmoids, flare ribbons with J-shaped ends and changes in the 3D configuration of the magnetic field before and after the eruptions \citep{2012A&A...543A.110A,2013A&A...549A..66A,2013A&A...555A..77J}. The process described by the standard model requires a triggering mechanism, which will produce an unstable configuration, and drivers which will initiate the eruption \citep[see][for a comprehensive review of the theoretical descriptions that have been put forward so far]{Green18}.

In this context, two quantities are of interest regarding CME initiation, the magnetic helicity and the decay index. The magnetic helicity \citep{1984JFM...147..133B} is a measure of the degree of linkage, twist, and writhe of the magnetic field lines and accumulates as the active regions emerge and build up free magnetic energy \citep[see e.g.][]{2012ApJ...759L...4T,2013ApJ...772..115T}. As a quantity, the magnetic helicity is conserved during dissipative processes but it is removed from active regions during eruptions. Its properties make it crucial in understanding the triggering mechanisms of CMEs as there are indications that when certain thresholds of free magnetic energy and helicity are exceeded, an eruption is initiated \citep[e.g., see discussion in][]{georgoulis_2019}. Thus, proxies of the instantaneous budget and accumulation rate of magnetic helicity have received attention as possible predictors of eruptions \citep{2010ApJ...718...43P,park12,Georgoulis_2013,2021JSWSC..11...39G}. On the other hand, the decay index, which is derived from magnetic field extrapolations, describes how fast the constraining magnetic field decays and, thus, determines whether a full eruption could take place or not \citep{2005ApJ...630L..97T,2008ApJ...679L.151L,2014ApJ...785...88Z,2015ApJ...814..126Z}. A faster decrease, i.e., higher decay index, reflects more favourable conditions for eruption. Based on the aforementioned works, three predictors based on the decay index were developed by the FLARECAST consortium \citep{2021JSWSC..11...39G}, namely the mean value of the decay index above PILs, the height at which the decay index reaches 1.5 and the ratio between this height and the length of the PIL.

From the studies already mentioned in Section~\ref{mag_params}, the ones by \citet{2006ApJ...644.1258F} and \citet{2008ApJ...689.1433F} have specifically addressed the prediction of CMEs. They highlighted the importance of twist and free energy in CME-productive active regions and their non-potentiality as represented by the $WL_{SG}$, that is, the gradient-weighted length of the neutral lines in an active region. More recently, \citet{2018ApJ...860...58V} demonstrated how calculating known parameters along manually tracked PILs can improve the correlation between magnetic parameters and CME characteristics.

The issue of determining which magnetic parameters are more suitable for CME prediction was tackled through machine learning by \citet{bobra_ilonidis16}, for a sample of SHARP data. In their work they discussed the distinction between extensive and intensive magnetic characteristics, and concluded that the latter are the most important for CME prediction. Extensive characteristics are the ones that scale with the size of the active region, such as the total magnetic flux, the total electric current density, and other ``total'' quantities. On the other hand, parameters that do not scale with the active region size, such as average quantities or quantities relevant to specific source region features such as PILs, etc. are non-extensive or intensive. In this context, quantities like, $B_{eff}$, $logR$, $WL_{SG}$, but also $I_{NN,tot}$ are intensive, since they correspondingly emphasize the presence of PILs, represent the magnetic flux in their vicinity, quantify their length or are proxies of the net electric currents injected to the corona.

Aiming to further explore the relevance of such PIL-related quantities, \citet{kontogiannis19} gathered a sample of 32 active regions/events of Solar Cycle 24 and calculated ten predictors, including some widely used ($R$,$WL_{SG}$,$B_{eff}$) and recently introduced ones ($I_{NN,tot}$,$I_{NN,max}$,$E_{Ising}$,$E_{Ising,part}$,$E_{Ising,spot}$, the length of PIL and the associated magnetic flux). The predictors were calculated along an interval of 24\,h before eruption and their daily-averaged values were correlated with kinematic characteristics of eruptions. It was found that the total unsigned non-neutralized electric currents and the length of the PIL were the ones most strongly correlated with CME characteristics, whose correlations improved when only the fastest events were considered. Their results were largely in line with earlier studies \citep{2018ApJ...865....4P}, but also demonstrated the relationship between net electric currents and strongly sheared PILs as well as their potential use in CME prediction. 

Lacking detailed measurements of the coronal magnetic field vector, some new parameters have attempted to include information relevant to the magnetic field in higher atmospheric layers using data-driven modelling. \citet{2019ApJ...883..112P} derived the parameters $\zeta$ from a data-driven magnetofrictional model, which incorporates information on the existence of magnetic flux-ropes and the Lorentz force. The eruption metric $\zeta$ is space- and time-dependent and can give information on the location of the eruption, while its maximum value, averaged over the entire time series of magnetograms can be used to distinguish eruptive from non-eruptive regions. In a follow-up study \citep{2019ApJ...886...81P} they elaborated their method and produced the quantity $\Lambda$, which they derived from the distribution of $\zeta$, as a more suitable parameter for operational forecasting of CMEs. \citet{2020ApJ...894...20L} applied non-linear force-free field (NLFFF) extrapolation to determine the twist of the magnetic field lines and discriminate between the core and the ambient magnetic fields. They produced the parameter $r_{m}$, i.e., the ratio of the magnetic flux associated with high twist to that of the overlying magnetic flux. However, they reported a moderate predictive potential of the proposed quantity (a relative measure of the core and constraining magnetic field), because further information from flare ribbons was required to pinpoint flare locations. \citet{2022ApJ...926L..14L} calculated the twist in regions close to the PIL and then divided the twist parameters with the total magnetic flux of the active region, which they consider representative of the constraining magnetic field. They concluded that the new parameter could distinguish effectively, for the examined sample, the eruptive from the non-eruptive cases.


\section{Quantifying the temporal evolution of active regions}
\label{time_series}

In the vast majority of the studies discussed so far, the formulated magnetic parameters were encapsulating a point-in-time information and thus represented a static condition in the evolution of active regions. However, active regions evolve dynamically and it is exactly the nature of this evolution that leads to major events \citep{georgoulis_2019}. The quantities used as predictors offer the chance to study the evolution of active regions in terms of their non-potentiality, revealing evolutionary patterns and mechanisms in action. Indeed, several exploratory studies on the development of non-potentiality in eruptive active regions have reported specific pre-eruptive patterns. 

The systematic provision of good quality LOS magnetograms by MDI facilitated the study of the helicity injection rate in flaring active regions \citep[see e.g.,]{2002ApJ...577..501K,2007ApJ...671..955L}, which showed that major flares followed specific threshold values of helicity and long intervals of increase. \citet{2008ApJ...686.1397P} found that before 11 X-class flares the magnetic helicity underwent a two-phase evolution: a monotonic accumulation, followed by a constant phase. The rate of helicity accumulation was more strongly correlated with the soft X-ray flux than the amount of helicity itself. Although the rate of helicity injection itself was not sufficient to describe the helicity history of an active region, flaring regions exhibited statistically twice as high helicity injection rate \citep{2010ApJ...718...43P}. The pre-eruptive behavior of NOAA 11158 was studied by \citet{2013ApJ...772..115T} in terms of free energy and magnetic helicity. \citet{2013Entrp..15.5022G} further studied the evolution of the same region in terms of the effective connected magnetic field strength, $B_{eff}$, and compared it with that of multi-scale parameters. The pre-eruptive evolution of the horizontal gradient proxies introduced by \citet{korsos14,korsos15} were studied for a few active regions indicating a specific evolutionary pattern before major flares. Other non-potentiality metrics such as the non-neutralized electric currents and the Ising energy exhibited structure with distinct peaks before major flares \citep{kontogiannis17,kontogiannis18}. \citet{kontogiannis19} showed that, within the 24 hours that preceded eruptions, different parameters exhibited different temporal evolution; the correlation of the parameters values with the characteristics of CMEs could increase if, instead of daily averages one considered instantaneous values on specific times before the events. Recently, \citet{2020ApJ...897L..23K} examined, for a limited sample of $\delta$-spot active regions, the periodicities of the magnetic helicity injection rate, which they decomposed into the emergence and the shearing term. They noticed that the most flare-productive ones shared a common periodic behavior. \citet{2022ApJ...925..129S} extended the analysis to a balanced set of 28 active regions, and reported that the emergence term of the magnetic helicity injection rate in flaring regions (i.e., those that hosted X-class flares) exhibited mostly shorter period oscillations (shorter than 10\,h) a few hours prior to X-class flares. However, in both studies no discussion of the origin of these oscillations was provided. This non-exhaustive account of studies and results attests to the wealth of information contained in the temporal evolution of the magnetic characteristics of active regions. Recent studies have started to incorporate them into flare prediction, utilizing the almost uninterrupted provision of photospheric magnetograms from HMI for more than one solar cycle.

\begin{figure*}
\centering
\includegraphics[width=16cm]{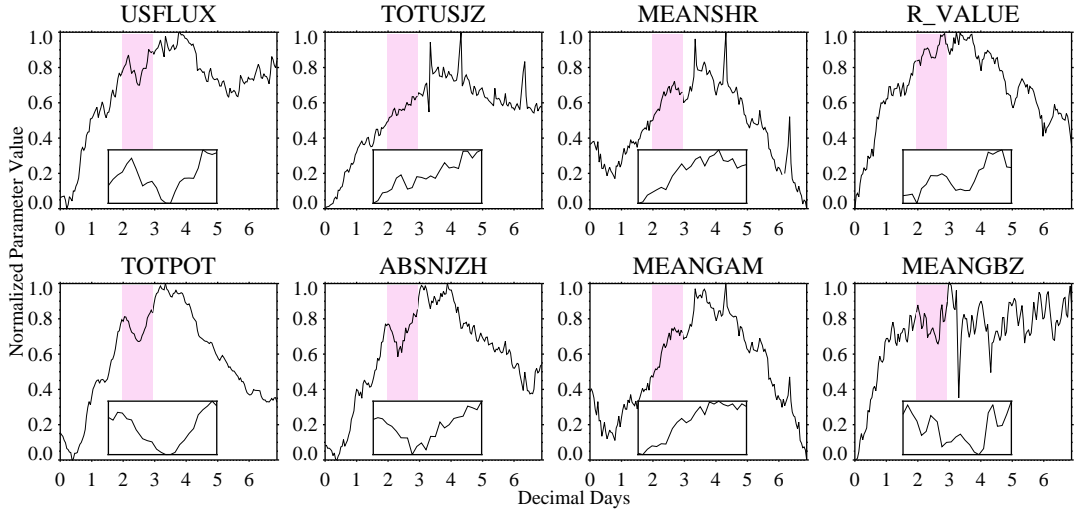}
\caption{Temporal evolution of eight SHARP parameters for active region NOAA 11429. From top left to right bottom, these parameters are the total unsigned magnetic flux, the total unsigned vertical electric current density, the shear angle, Schrijver's $logR$, the total photospheric magnetic free energy density proxy, the absolute value of
the net current helicity, the mean angle of field from radial direction, and the mean horizontal gradient of the vertical magnetic field. The insets show the evolution of the same parameters during the 24\,h before the twin X-class flares of 7 March 2012 (shaded with lilac in the original plots). All parameters are normalized to their minimum and maximum values, as in previous figures. }
\label{fig:fig4}
\end{figure*}

The evolution of eight SHARP parameters for active region NOAA 11429 is shown in Figure~\ref{fig:fig4}. As also discussed for NOAA 11158 (Figure~\ref{fig:fig2b}), these parameters follow a generally similar trend, but with differences regarding their overall steepness of increase and the existence of distinct peaks. Additionally, as demonstrated by \citet{kontogiannis18} different parameters have different sensitivity on instrumental and geometric effects. Focusing on the evolution of these parameters during the day before the X-class flare of 7 March 2012 (insets in the panels of Figure~\ref{fig:fig4}, it is well seen that some parameters exhibit an almost monotonic increase until the onset of the event, while others show clear peaks and valleys. While this is only one example, it is reasonable to inquire whether specific parameters exhibit a systematic behavior, how this behavior can be quantified and what is its importance to flare and CME prediction. 

Before reviewing how temporal information of magnetic parameters has been incorporated to predictions, it is worth noting the that the flaring history of active regions itself has been shown to be a good indicator of future activity. \citet{2005SpWea...3.7003W} described a methodology wherein the probabilities of future GOES events (particularly strong ones) can be inferred using the distributions of prior ones. The flaring history is routinely included as an extra feature in explored and operational forecasting schemes and is often found to be one of the top performers \citep{2014SpWea..12..306F,2019ApJ...877..121L,leka_etal19a,leka_etal19b,park_etal20}.

Regarding dynamic effects, so far these have been incorporated into predictions in two ways. One of them is to use entire time series of predictors to train machine learning models. In this case the temporal information already contained in the time series is utilized, without this having been quantified with specific parameters \citep[see e.g.,][]{2010SoPh..263..175H,2010ApJ...709..321Y,2010ApJ...710..869Y,2019SpWea..17.1404C,2020ApJ...890...12L}. The other strategy is to extract features related to the history of active regions, either their past flaring activity or the trend exhibited by their complexity parameterizations. Usually, this requires the choice of time intervals, over which the statistics of the flare history will be calculated or features like derivatives or summary characteristics will be extracted.  

\citet{2012SoPh..281..639L} included also the change in the white-light area, as an indicator of flux emergence or cancellation. They did so by complementing the McIntosh classes with three additional categories, depending on the direction of area change, namely ``decrease'', ``steady'' and ``increase''. They noted that higher probability rates were associated with increasing area. \citet{2018JSWSC...8A..34M} proposed a forecasting scheme based on the evolution of complexity of active regions. They did so by calculating the flaring rates associated with the change of McIntosh classes over 24\,h intervals. Evolution towards class of higher complexity was associated with higher flaring rates and validation and bias correction for the differences across solar cycles indicated an improved performance for the forecast based on the McIntosh evolution.

Aiming to include measures of the existence and intensity of magnetic flux emergence \citet{2015A&A...579A..64A} introduced nine flux-emergence-related features. To calculate these features they subtracted subsequent magnetograms and isolated the regions that stood above a 3$\sigma$ difference level. From this they calculated the sum (net and absolute), the area and the statistics (mean, standard deviation, minimum and maximum over the mask). Additionally, they used this 3$\sigma$ mask to calculate the gradient of the second image. Their feature space, consisting both of static (magnetic and fractal-related) and temporal (flux-emergence-related) parameters was then used to train Regression Vector Machines and, in a follow-up study, Support Vector Regression models \citep{2015ApJ...812...51B}. Their results support the overall discriminatory power of the predictor ensemble to classify flaring and non-flaring events but no conclusion on the importance of the temporal parameters was drawn. \citet{2017ApJ...835..156N} trained three ML algorithms using features extracted from magnetograms and EUV images of active regions. These features included also temporal derivatives of features and the flaring history, as recorded by GOES. The latter was deemed an important feature that improved flare prediction in their model. Although the point-in-time values of magnetic parameters were found overall more important than their differentials, concerning the latter, the derivatives calculated over 24\,h were found more important than those over shorter intervals.

\citet{2018JSWSC...8A..25L} described the Discriminant Analysis Flare Forecasting System (DAFFS) of the North Western Research Associates (NWRA). Along with the standard predictors introduced in their earlier work, the scheme included also previous flaring information and magnetic parameter time series information. The latter was introduced by modelling 6 hour-long time-series' segments with a linear function. The time-series segments end on the time of the issuing of the forecast. The derived slope and intercept were treated as predictors. Although the previous flaring information is among the top performing predictors, the inclusion of the preceding temporal variation of the magnetic parameters in the form of slope and intercept did not lead to a notable improvement in the forecasts, even though they could contribute to equally competent predictions. It should be noted in passing that various combinations of the many parameters utilized by \citet{2018JSWSC...8A..25L} performed similarly well, within the error margins. 

\citet{2018SoPh..293..159L} used a Detrended Fluctuation Analysis to remove persistent patterns and quantify the random variations in consecutive non-overlapping segments of time series. The detrended fluctuation and the time series averages were positively correlated with the imminent 8-day flare index. In a follow-up study \citet{2020SoPh..295..123L} found that flaring regions exhibited higher values of the scaling exponent of the detrended fluctuations. \citet{2019RNAAS...3..157P} used spline interpolation on segments of time series, and then utilized the spline coefficients as additional predictors, along with standard ones, which lead to a slight improvement of predictions.

Notwithstanding different strategies and approaches, and despite still lacking a definitive assessment on the improvement of predictions when incorporating time series, their importance is generally accepted. Current facilities/instruments not only provide ample data but also facilitate the construction of standardized multivariate time series datasets for further studies. \citet{angryk_etal20} created such a benchmark data product, which comprises time series of selected magnetic properties and, most importantly, an ``error-free'' association with flare occurrences. Data sets like this provide a test bed to compare different approaches, not only for prediction models based on time series but also for those relying on point-in-time information. 

A more in-depth exploration of characteristic patterns in large statistical samples is still missing and such a task could benefit both exploratory and operational research. Time-series based operational forecasting of flares and CMEs can not be realized without standardized, continuous observations of the Sun. Since it takes at least a few days of observations to produce a times series of magnetic parameters, the current instrumentation limits forecasts only for the regions closer to the central meridian. Therefore, the capabilities of incorporating time series in forecasts can be fully exploited in the future if continuous observations from different vantage points complement the existing ones, offering access to active regions before they rotate into the earth-facing side of the Sun \citep{2015SpWea..13..197V}.

\section{Summary and outlook}
\label{summary}

The increased awareness of national/international space agencies and stakeholders on the adverse consequences of space weather \citep{schrijver_etal15} has fueled fundamental and operational research in the relevant fields of heliophysics. This fact reflects on the abundance of studies on space weather prediction during the past decade. Before and up to the 1990's studies were limited by the limited available observations; during the past two decades a multitude of instruments provided access to many facets of solar activity, from the emergence of new magnetic flux on the solar surface to influences seen on outer corona, solar wind and interplanetary space. Nowadays, at least in solar physics, it is perhaps hard to imagine how research would be without the standard flow of continuous, high-quality magnetic field measurements and coronal imaging provided by missions like SDO. 

Consequently, flare and CME prediction has moved forward by creating an inventory of active region characteristics and prediction methods. The characteristics (predictors or parameters) amass to more than two hundred \citep[see e.g.,][]{2021JSWSC..11...39G} and incorporate quantities relevant mostly to the photospheric magnetic field, but highlight different aspects, such as size, various physical quantities (electric current, magnetic energy, magnetic helicity) and proxies thereof, PIL characteristics and connectivity. Regarding the prediction methods, these are now dominated by ML, with increasing sophistication on sample selection and validation. The community is working intensely towards comparing methods and setting standards, which will help provide the most accurate possible forecasts and address the needs of interested parties \citep[see e.g.,][]{barnes16,leka_etal19a,leka_etal19b,park_etal20}.

Notwithstanding the access to large, high-quality statistical samples and sophisticated methods, the prediction of solar eruptions remains probabilistic, due to the nature of solar eruptions and the limitations of observations. Flares are inherently stochastic, in the sense that given favourable conditions, they may happen at any time and the available free energy can be channeled in any combination of flare magnitudes and CME energetics, producing one major event, several smaller ones or both. Furthermore, the characteristics of flare- and CME-productive active regions are relevant to the photosphere, despite the fact that the magnetic interactions that power solar eruptions take place $\gtrsim$2\,Mm higher. It is, therefore, reasonable that the magnetic parameters exhibit a degree of redundancy and that there is also no conclusive answer on which single parameter can provide the best results. This has been already highlighted by \citet{2007ApJ...656.1173L}, who showed that the inclusion of many properties does not necessarily improve the forecast, and more recently by \citet{campi19}, who also concluded that the inherent stochasticity of flares does not allow a binary (i.e. ``yes/no'') forecast.  

On the other hand, it is largely supported that probabilistic forecasts can improve when using better data sources and methods \citep{leka_etal19a}. Since no systematic provision of coronal magnetic field observations is expected in the immediate future, one course of action is to continue searching for improved predictors based on the data provided by current instrumentation. The studies cited in this review show how results from fundamental research can be transformed into promising new predictors, corroborating that there is indeed still room for improvement in quantifying aspects of imminent eruptive activity. New properties now incorporate more sophisticated information from image processing, photospheric flow fields, spectroscopy and data-driven modelling, the last two of which are, at the moment, our only way to probe higher atmospheric layers. On the one hand, spectra and observables from transition region and the corona provide information more relevant to the atmospheric layers where the initiation of flares and CMEs takes place, with promising future implications \citep{2021ApJ...917...27J}. On the other hand, data-driven modelling, although it cannot fully compensate for our lack of coronal magnetic field measurements, is our only way to probe the three-dimensional vector magnetic field in the solar atmosphere and assess the competing roles of the core and ambient magnetic field in CME initiation \citep{2018Natur.554..211A}. Further work is required here to instill this information into more refined parameters and advance CME prediction.

Finally, there is a growing body of work on exploiting the dynamic behavior of active regions, utilizing the, now abundant, time series of existing magnetic properties. Moving from the ``static'' point-in-time treatment of source regions to monitoring their entire evolution from emergence to decay is a huge leap forward for the relevant fields of research. Some of this dynamic information is incorporated into prediction schemes. Nowadays ML offers a huge variety of methods to treat time-series, for knowledge discovery, classification and/or prediction \citep{time_ser_bakeoff} and it is up to the community to proceed towards this direction in a more systematic way, especially given the availability of openly accessible, standardized datasets \citep{angryk_etal20}. 

To the authors experience, a common question is how new missions, like the Solar Orbiter \citep{2020A&A...642A...1M} and the new generation of ground-based solar telescopes, such as the Daniel K. Inouye Solar Telescope \citep[DKIST;][]{2020SoPh..295..172R} and the European Solar Telescope \citep[EST;][]{2022arXiv220710905Q}, can contribute to the task of flare and CME prediction. A short response would be that they cannot, since statistically significant samples of standardized, continuous observations are necessary to transform any systematic behavior into parameters suitable for prediction and to test their potential efficiency. They are, however, expected to contribute indirectly by enhancing our understanding of solar eruptive phenomena and pointing to specific connections between solar drivers and space weather effects. This was the case with Hinode and IRIS missions. Although they were not intended for operational space research they did contribute to characterizing the potential of active regions to erupt, either by igniting further research \citep{2012ApJ...759....1G,2013ApJ...774..122H,2016A&A...588A..16S} or, eventually, by directly producing sizeable samples of observations \citep{2020ApJ...891...17P}. Therefore, it is expected that those new facilities and instrumentation will stimulate further fundamental research and open new windows in solar observation, particularly by supplying accurate measurements of the magnetic field at higher atmospheric layers. 

A final note is reserved here for the important role of big projects like FLARECAST \citep{2021JSWSC..11...39G}. The efficient prediction of flares, CMEs and other space weather aspects rely on a gamut of techniques and (sub-)fields. These include solar and space physics, computer science, and machine learning, complemented by outreach, which is necessary to coordinate communication with interested parties, raise awareness and stimulate citizen science projects. Therefore, critical elements for such a task are interdisciplinarity and collaboration between communities and stakeholders. FLARECAST brought together experts from these areas to produce a novel flare-prediction service. By gathering all available parameters and implementing new ideas, not only it created a comprehensive set of magnetic parameters but also enriched it with new, promising ones. It is expected that such an accomplishment will serve as a prototype for future endeavours and also ignite new ways to characterize the flaring and eruptive potential of solar active regions.

\section{Acknowledgments} 
I would like to thank the two anonymous reviewers for providing detailed and constructive comments, which significantly improved the content and presentation of this paper. I would also like to thank the Scientific Organizing Committee of the $16^{th}$ European Solar Physics Meeting (ESPM-16), for the invitation to present a review on the topic presented in this paper. I.Kontogiannis is supported by KO\,6283/2-1 of the Deutsche Forschungsgemeinschaft (DFG). For this paper, data from SDO/AIA, SDO/HMI and particularly SHARP were used. These are courtesy of NASA/SDO and the AIA, EVE, and HMI science teams and are publicly available through the Joint Science Operations Center at the \url{jsoc.stanford.edu}. Part of the calculations was performed using source codes made publicly available by the H2020 project FLARECAST (\url{https://dev.flarecast.eu/stash/projects/}).

\bibliography{sample}

\end{document}